\documentclass{elsart}
\usepackage[latin9]{inputenc}
\usepackage{verbatim}
\usepackage{amsmath}
\usepackage{amssymb}

\providecommand{\tabularnewline}{\\}

 \usepackage{graphicx}

\usepackage{subfigure}

\renewcommand{\thetable}{\Roman{table}}
\newcommand{\sigv}{{\bs{\sigma}}}
\newcommand{\rhov}{{\bs{\rho}}}
\newcommand{\bs}[1]{\boldsymbol{#1}}
\makeatother

\begin{document}

\begin{frontmatter}



\title{Characterising heavy-tailed networks~using q-generalised entropy
and q-adjacency kernels}

\author{Ismo T. Koponen\corauthref{cor}},
\corauth[cor]{Corresponding author}
\ead{ismo.koponen@helsinki.fi}
\author{Elina Palmgren},
\author{Esko Keski-Vakkuri}

\address{Department of Physics, P.O. Box 64,
FI-00014 University of Helsinki, Finland}

\begin{abstract}
Heavy-tailed networks, which have degree distributions characterised
by slower than exponentially bounded tails, are common in many different
situations. Some interesting cases, where heavy tails are characterised
by inverse powers $\lambda$ in the range $1<\lambda<2,$ arise for
associative knowledge networks, and semantic and linguistic networks.
In these cases, the differences between the networks are often delicate,
calling for robust methods to characterise the differences. Here,
we introduce a method for comparing networks using a density matrix
based on q-generalised adjacency matrix kernels. It is shown that
comparison of networks can then be performed using the q-generalised
Kullback-Leibler divergence. In addition, the q-generalised divergence
can be interpreted as a q-generalised free energy, which enables the
thermodynamic-like macroscopic description of the heavy-tailed networks.
The viability of the q-generalised adjacency kernels and the thermodynamic-like
description in characterisation of complex networks is demonstrated using a simulated set of networks, which
are modular and heavy-tailed with a degree distribution of inverse
power law in the range $1<\lambda<2$.
\end{abstract}

\begin{keyword}
Heavy tailed networks \sep adjacency kernels \sep q-entropy \sep generalised thermodynamics 
\end{keyword}

\end{frontmatter}

\clearpage{}

\section{Introduction}

When associative knowledge or term associations between words and
terms are represented as networks, the networks are often found to
be heavy-tailed \cite{Koponen:2018b,Lommi:2019,Morais:2013}. Being
heavy-tailed means here that the trailing edge of distribution of
low frequency for node degrees $d$ is not exponentially bounded but
instead exhibits an inverse power-law type of decay $P(d)\propto d^{-\lambda}$
with $\lambda \in\,]1,3]$. In addition, such networks may also have
a clear modular or community structure, originating, for example,
from the thematic content of nodes \cite{Interdonato:2019}. For example,
networks representing students' associative knowledge have distributions
of degree and Katz centralities which are heavy-tailed and can be
fitted with inverse power laws with exponents in the range of 1 to
2 \cite{Koponen:2018b,Lommi:2019}. The heavy tails with such values
of $\lambda$ also appear to be ubiquitous in word frequency, ranking
distributions and many linguistic networks \cite{Morais:2013,Thompson:2014}.
Powers from 1.5 to 2 are encountered in information and knowledge
networks like Wikipedia and in semantic networks resulting from acquiring
knowledge (terms) from large knowledge networks \cite{Thompson:2014,Masucci:2011}.

The research on associative networks often focuses on finding individual
key items in the networks and their connectivities. Communicability
centrality, as introduced by Estrada, and the closely related Katz centrality
have been useful in exploring the connectivities of the nodes through
long paths (or walks) \cite{Koponen:2018b,Lommi:2019}. 
In practice, the real networks in such applications are always quite small (about 1000-1500 nodes) and centrality measures 
have usually very large statistical variation. Therefore, it would be advantageous to seek for a descriptions which use all available information of the network states. 
However, taking
the networks as holistic systems, that is, as connected sets of items,
requires different approaches based on the description of the networks
on macro-level \cite{Biamonte:2019,Domenico:2016}. For this, matrix
transformations based on a complete adjacency matrix that describes
the network are a promising starting point. For example, the communicability
centrality is related to a matrix transformation which is obtained
as an exponential transformation, while Katz centrality is related
to Neumann transformation \cite{Kunegis:2013, Benzi:2015}. Here, we introduce
a q-generalised \cite{Tsallis:1988,Borges:1998,Yamano:2002} matrix
transformation (called a q-adjacency kernel in what follows), which
interpolates between these two well-known transformations. It is shown
that the q-adjacency kernel provides a starting point for a thermodynamic-like
macroscopic description of heavy-tailed networks and allows us to define a quantity
that can be interpreted as the q-generalised free energy of a network.

The q-generalised free energy for a heavy-tailed network is obtained here by a derivation based on q-generalised information theoretic entropy. The information theoretic entropic measures have recently found many application in the characterisation of complex networks, multilayer and multiplex networks \cite{Biamonte:2019, Domenico:2016,Re:2014, Bagrow:2019}. For example, the Jensen-Shannon divergence, which can be seen as a symmetrised and regularised form of the Kullback-Leibler divergence, has proved to be a flexible and robust information theoretic entropic measure for comparison of the networks, because it is symmetric, bounded and it provides connection points to several related divergence 
measures within statistical physics \cite{Re:2014, Bagrow:2019}. The quantum version of the Jensen-Shannon divergence can be utilised effectively to measure the similarity between networks, as based on quantum walks \cite{Minello:2019}.  Another class of information theoretic entropies is provided by R\'enyi and Tsallis entropies \cite{Domenico:2016,Tsallis:1988,Muller:2013,Tsekouras:2005} which are closely related classes of entropies. These entropies belong to a class of q-generalised  information theoretic measures, which have recently found applications in many complex systems, quantum systems and entangled systems \cite{Domenico:2016}. The R\'enyi and Tsallis entropies form a family of information theoretic measures that generalise the Shannon entropy, and can be used as measures of mutual information and relative entropies (see e.g. \cite{Muller:2013}). Regarding the characterisation of heavy-tailed networks, the generalised q-entropies have the advantage of providing a parameter (q-index) to tune for specific properties of heavy-tailed networks, either for the low or high frequency part of the distributions, thus allowing foraging for information of the desired properties of the distribution \cite{Gerlach:2016, Dias:2018, Altmann:2017}.  Consequently, the Tsallis and R\'enyi entropies have found applications in linguistic and semantic networks \cite{Gerlach:2016} as well as in comparisons of texts based on word frequency occurrence \cite{Dias:2018, Altmann:2017}, all these are cases where heavy-tailed networks are of interest.  Here, we show that the Tsallis entropy and the q-generalised Kullback-Leibler (termed Kullback-Leibler-Tsallis in what follows) divergence \cite{Abe:2003b,Abe:2004} (see also \cite{Martins:2009,Furuichi:2014}) are particularly suitable as a starting point to derive a q-generalised free energy to characterise the heavy-tailed network, thus providing access to a macro-level, thermodynamic-like description of the heavy-tailed networks. 

To test the applicability of q-adjacency kernels for networks that
are heavy-tailed and modular, such as associative knowledge networks,
we use a recently introduced generative model to produce networks
that resemble the empirical networks \cite{Koponen:2018c}. The minimal model to generate the networks is introduced first, in section 2.
In section 3 we discuss in detail how q-generalised adjacency kernels are obtained, and how the thermodynamic-like interpretation is based on q-generalised Kullback-Leibler-Tsallis divergence. The section 3 ends with a discussion on
how by using ensembles of the simulated heavy-tailed networks the validity of the derived theoretical results can be tested. Section 4 provides the results showing that the heavy-tailed networks yield to  the thermodynamic-like macro-scale characterisation, based on the q-generalised entropy and free energy. In addition, Appendix A contains a sequel result of practical utility, for comparison of heavy-tailed networks. Finally, the implications and the practical use of the results are discussed.

\section{Minimal generative model and simulations}

Many real networks have hub-like nodes so that their degree distributions can be characterised as heavy-tailed distributions. Here, heavy tails
refer to a form of the trailing edge of a distribution that decays considerably more slowly than a normal distribution, or to exponential
distributions for which the trailing edge resembles to some degree the inverse power-law type distribution given as 
\begin{equation}
P(d)\propto d^{-\lambda'},\:\:\:\lambda'\in\,]1,3].\label{Peq1}
\end{equation}
The inverse power-law type distributions are used here as a model of heavy-tailed distributions. However, it is not assumed that the
heavy-tailed distributions are genuinely inverse power-law or can be strictly characterised as such, since this rarely seems to be the case \cite{Broido:2019}. 
Nevertheless, for practical reasons, it is useful to consider the networks with heavy-tailed distributions using the model of inverse power-laws, because many essential characteristics
of the networks are captured by that class of distributions \cite{Holme:2019}.
In addition to heavy tails, many real networks are also highly modular, with the modularity emerging e.g. from content based features \cite{Interdonato:2019}
or thematisation \cite{Lommi:2019}. Here, we focus on empirically interesting cases of associative knowledge networks with $\lambda'\in\,]1,2]$ with high modularity $Q$ (as defined following Newman \cite{Newman:2004}) 
within range $0.8<Q<0.9$ \cite{Koponen:2018b,Lommi:2019}.

We utilise here the generative model to produce a set of heavy-tailed networks that have a highly modular structure
\cite{Koponen:2018c}. The model is based on generation of affinities for a fixed set of nodes, with the nodes assigned to different,
pre-fixed classes. Here, we use parameters which generate networks of approximately N=1000 connected nodes and with inverse powers $\lambda$ of
degree-distributions obtaining values $\lambda$ =1.3, 1.5, 1.7 and 1.9. The networks are generated using, for affinities $\pi_{k}$ of nodes,
the distribution 
\begin{equation}
P(\pi_{k})=P_{0}\;[1-(\lambda-1)\,\Lambda\pi_{k}]^{-1/(\lambda-1)},\;\;\;\lambda\in\,]1,2],\;\;\;\Lambda>0.\label{eq2}
\end{equation}
It is interesting to note that Eq. (2) can also be written in the form of a q-exponential $P(\pi_{k})/P_{0}={\rm exp}_{q}[\Lambda\pi_{k}]$,
which is the q-generalisation of the exponential function \cite{Tsallis:1988,Borges:1998,Yamano:2002}.
The parameter $\lambda$ determines the inverse power of the resulting degree distribution of the network, while $\Lambda$ controls the cut-off
of affinities, and $\pi_{k}<\Lambda(\lambda-1)$. The affinity distribution can be derived in several ways and by using several parametrisations,
\cite{Koponen:2018c,Servedio:2004,Caldarelli:2002}. What is essential here is that the affinity distribution in Eq. (2) can be used to generate
networks with the desired inverse power. Due to modularity, however, the resulting inverse power of degree distribution is never exactly
$\lambda$ but usually somewhat larger, $\lambda'>\lambda$ \cite{Koponen:2018c}.

In simulations, only one nested modular structure is used, which is three-tiered; the smallest modules contain N' potential nodes, the second
tier three of these smaller modules and thus 3xN' , and the highest and third tier contains three of the modules of 3xN' nodes, thus forming a set of N=3x(3xN') nodes in total.
The connections between N nodes are established probabilistically as governed by the affinity distribution. Therefore, not all nodes in the nested tiers will be connected. 
With parameters chosen so that the resulting network has the size of about N=1000 connected nodes,  the inverse power in range $1.3 < \lambda < 2.0 $ and the modularity in range $0.8< Q <0.9$. 
The modularity of the networks is relaxed step-by-step, by rewiring until no changes due to rewiring are observed. In the rewiring, the degree sequence of the nodes is preserved.
We have shown that such affinity-based linking of nodes with minimal other assumptions reproduces the empirically discovered properties
of associative knowledge networks quite adequately \cite{Koponen:2018c}.

The simulations to generate networks and their analysis are carried out in \texttt{Mathematica} using the \texttt{IGraph} package \cite{Csardi:2006}.
\texttt{IGraph} provides functionality for generating efficiently affinity-based networks simply by providing the probabilities $\pi_{k}$
for the routine \texttt{IGStaticFittnessGame}. The output of the routine is a network with a predetermined number of links, linked according
to the probabilities $\pi_{k}$ drawn from distribution $P(\pi_{k})$ in Eq. (2). The relaxation of modularity by rewiring is done using
routine \texttt{IGRewire}. For each network, the simulations provide the adjacency matrix ${\bf A}$, which is then used to obtain the
desired adjacency kernels to characterise the networks. Next, we turn to how the adjacency kernels are constructed and how the heavy-tailed
networks can be characterised using them.

\section{Thermodynamic-like description of networks with heavy tails}

The thermodynamic-like description of the networks with heavy tails can be constructed using a suitable generalisation of the matrix transforms that have
been used more traditionally to describe and characterise networks. It is shown how these generalisations provide a basis for constructing
the thermodynamics-like description and and generalised free energy -like quantity for characterisation of the network. In that, the discussion closely follows the study
by Abe and Rajagopal \cite{Abe:2003}, who derived the q-generalised thermodynamics for q-generalised ensembles. The motivation of \cite{Abe:2003} was to develop a non-extensive
version of thermodynamic laws for finite systems where the infinite volume thermodynamic limit is not applicable 
and surface effects violate extensivity. Similar constraints apply to finite complex networks,
hence we are interested in exploring connections to non-extensive thermodynamics by applying the results and approaches of \cite{Abe:2003}
to networks with heavy tails.

\subsection{Adjacency kernels and q-kernels}

On the most basic level, networks are described by the set of nodes (vertices) and links (edges) between the nodes. This information is
provided by the network's adjacency matrix ${\bf A}$, whose element $a_{ij}=[{\bf A}]_{ij}$ has the value 1 if the nodes $i$ and $j$
are connected. Otherwise its value is 0. From the adjacency matrix, many other relevant properties of the network become available through
matrix transformations. Following previous studies on adjacency kernels \cite{Kunegis:2013,Benzi:2015} we define a graph kernel as a symmetric and positive semidefinite
function which denotes a certain similarity or proximity between two nodes in the graph. All such graph kernels of interest here are based
on the adjacency matrix ${\bf A}$ of the graph and are thus called adjacency kernels. Two well-known adjacency kernels, the exponential
and the Neumann kernel, are related to path counting between nodes, thus providing information on how the nodes are globally connected
\cite{Kunegis:2013,Benzi:2015}.

The exponential kernel is one of the matrix transformations most widely
used to analyse networks. The exponential kernel is defined as \cite{Kunegis:2013,Benzi:2015,Estrada:2012,Estrada:2009,Estrada:2012b,Dehmer:2008,Dehmer:2011}
\begin{equation}
{\rm exp}[\beta{\bf A}]={\bf {I}+\frac{\beta^{{\rm 1}}{\bf A}^{{\rm 1}}}{{\rm 1}!}+\frac{\beta^{{\rm 2}}{\bf A}^{{\rm 2}}}{{\rm 2}!}+\frac{\beta^{{\rm 3}}{\bf A}^{{\rm 3}}}{{\rm 3}!}+\ldots,}
\end{equation}
where matrix power ${\bf A}^{k}$ counts paths (walks) of length $k$. The exponential adjacency kernel weights the paths according to their
length $k$ by a weight factor $\beta^{k}$ and inverse of the factorial $k!$. Therefore, connections with short paths gain more importance
than those with long paths. The diagonal elements of the exponential kernel are called the Estrada index and the so-called communicability
centrality is obtained as a row sum of the off-diagonal elements \cite{Estrada:2009}.
The exponential kernel and the associated measures (the Estrada index
and the communicability centrality) perform well in finding globally important nodes. In addition, the exponential kernel is computationally
very robust and stable, which is an advantage for analysis of complex networks \cite{Estrada:2012,Estrada:2009,Estrada:2012b,Benzi:2013}.

The Neumann kernel \cite{Kunegis:2013,Benzi:2015} is another well-known adjacency kernel with many applications in characterising complex networks \cite{Katz:1953,Sharkey:2017}.
The Neumann kernel is given by \cite{Kunegis:2013,Benzi:2015} 
\begin{equation}
[{\bf I}-\beta{\bf A}]^{-1}={\bf {I}+\beta^{{\rm 1}}{\bf A}^{{\rm 1}}+\beta^{{\rm 2}}{\bf A}^{{\rm 2}}+\beta^{{\rm 3}}{\bf A}^{{\rm 3}}+\ldots,}
\end{equation}
where the weight factor $\beta$ must be chosen so that $\beta^{-1}$ is larger than the largest eigenvalue of matrix ${\bf A}$. The row
sum of the off-diagonal elements of the $i$th row in the Neumann kernel provides the Katz centrality \cite{Katz:1953,Sharkey:2017} of node
$i$, widely applied in social network analysis in finding the influential nodes \cite{Borgatti:2005}. The Katz centrality has also been successfully
used in finding the central key nodes in heavy-tailed associative networks, which are of interest here \cite{Lommi:2019}. One disadvantage
of the Katz centrality is that it is difficult to optimise the weight factor $\beta$ to maintain the stability of computation and the desired
resolving power \cite{Benzi:2015,Ghosh:2011}. On basis of effects of the choice of parameters in behaviour of the kernels and consequently, on ranking of nodes, it seems to be advisable
to strive for largest values of weight factor but to check how results depend on the choice of it \cite{Benzi:2015}.   

The generalisation of the walk counting, which interpolates between the exponential and Neumann kernels, is the q-generalised exponential kernel
\cite{Abe:2003} 
\begin{equation}
{\rm exp}_{q}[\beta{\bf A}]=[{\bf I}-(q-1)\beta{\bf A}]^{-1/(q-1)},\:\:\:q\in\,]1,2]\:\:{\rm and}\:\:\beta>0,
\label{Qkernel}
\end{equation}
where $\beta$ must be chosen so that $\beta<1/B$ with $B$ being the largest eigenvalue of matrix ${\bf A}$. The q-exponential adjacency
kernel can be expanded in series 
\begin{eqnarray}
{\rm exp}_{q}[\beta{\bf A}] & = & {\bf I}\!+\!\beta^{1}{\bf A}^{1}\!+\!\frac{q}{2!}\beta^{2}{\bf A}^{2}\!+\!\frac{2q^{2}\!-\!q}{3!}\beta^{3}{\bf A}^{3}+\!\ldots=\sum_{{\rm k}}c_{k}(q)\:\beta^{{\rm k}}{\bf A}^{{\rm k}}\\
\nonumber \\
c_{k}(q) & = & (1-q)^{k}\binom{\frac{1}{1\!-\!q}}{k}=\frac{(1-q)^{k}}{k!}\left(\frac{1}{1\!-\!q}+1-k\right)_{k}
\end{eqnarray}
where $\binom{\,\cdot\,}{\cdot}$ is the binomial coefficient and $(\cdot)_{k}$
is the Pochhammer ascending factorial \cite{Abramowitz:1972}. 
In the limit $q\rightarrow1$ the q-adjacency kernel approaches the exponential kernel since $c_{k}\rightarrow1/k!$ while for $q\rightarrow2$
it approaches the Neumann kernel when $c_{k}\rightarrow1$. The q-adjacency kernel interpolates between walk counting where walks of length $k$
are either weighted by $\beta^{k}$ (Neumann kernel) only or much more heavily with $\beta^{k}/k!$ (exponential kernel). The q-adjacency
kernel is also a positive semidefinite matrix and has non-negative eigenvalues when $\beta$ is smaller than the inverse of the largest
eigenvalue of the adjacency matrix. 
The q-adjacency kernel in Eq. (\ref{Qkernel}) opens up a path to two interesting generalisations.  First, the q-kernel provides a starting point for constructing the thermodynamic-like
description of the networks with a heavy-tailed degree distribution. Second, it leads to q-generalisations of the Estrada-type communicability centrality, which is of high practical utility in characterising heavy-tailed networks. However, while it is useful, we have not found a thermodynamic-like interpretation for it. Hence we have relegated the discussion of the generalised communacibility centrality to Appendix A. We proceed next to the details of the thermodynamic-like interpretation of the kernels. 

\subsection{Thermodynamic-like description of heavy-tailed networks}

The heavy-tailed networks generated by simulations and based on the generic model introduced in section 2 are described using adjacency
matrix \textbf{A}. Using the q-adjacency kernels we characterise the connectivity (or communicability) properties of the network by defining
the density matrix 
\begin{equation}
\boldsymbol{\rho}=Z_{q}^{-1}[{\bf I}-(q-1)\beta{\bf A}]^{-1/(q-1)} \equiv Z_q^{-1}\exp_q (\beta {\bf A}),\label{densityMatrix}
\end{equation}
where $Z_{q}={\rm Tr}[{\bf I}-(q-1)\beta{\bf A}]^{-1/(q-1)}$ is the
q-generalised partition function (compare with \cite{Abe:2003}). The density matrix provides a complete description of the network
and, in that sense, is a holistic measure for the characterisation of networks. Moreover, it can be taken as a starting point for robust
comparison of the structure and structural similarity of networks \cite{Biamonte:2019,Domenico:2016}.

\subsubsection{Thermodynamics corresponding to Gibbs states with $q=1$}

We review first a thermodynamic interpretation of the density matrix in the case $q=1$, where
\begin{equation}\label{Est}
\boldsymbol{\rho}=Z^{-1}\exp ( {\beta {\bf A}} )\ .
\end{equation}
There are two obvious ways to interpret (\ref{Est}) as a thermal (Gibbs) state. First, in a system with a Hamiltonian ${\bf H}=-{\bf A}$, 
\begin{equation}\label{Gibbs}
\boldsymbol{\rho}=Z^{-1}\exp ({-\beta {\bf H}}) \ .
\end{equation}
with a positive inverse temperature $\beta>0$, where the prefactor $Z=Z(\beta) ={\rm Tr}(e^{-\beta {\bf H}})$ is the partition function. For example, one could consider the Hamiltonian as the graph Laplacian ${\bf H}={\bf D}-{\bf A}$ without the diagonal matrix ${\bf D}$ of the node degrees (compare with \cite{Biamonte:2019,Domenico:2016}). Second, one can choose the adjacency matrix
directly as the Hamiltonian, ${\bf H'}={\bf A}$ (the prime is used to distinguish this alternative choice), but then one must choose a negative temperature $\beta' = -\beta<0$ to write $\exp (\beta {\bf A})=\exp (-\beta' {\bf H'})$ \cite{Estrada:2012,Estrada:2012b}. 
Negative temperatures arise naturally in systems with a bounded spectrum of energy eigenvalues (as in this case, {\bf H} has a finite set of eigenvalues).  
For convenience, we proceed with the first interpretation with the Hamiltonian ${\bf H}$.

The Gibbs state maximizes the von Neumann entropy 
\begin{equation}
\label{SvN}
S(\sigv) = -{\rm Tr}(\sigv \log \sigv) \,
\end{equation}
with the constraint that the expectation value of the Hamiltonian $\langle {\bf H}\rangle_{\sigv} = {\rm Tr}(\sigv {\bf H})$ is kept fixed. That is, with the constraint, the
density matrix $\sigv$ for which $S(\sigv)$ is maximised, is the Gibbs state $\sigv=\rhov$.
Substituting the Gibbs state to the von Neumann entropy, so that it becomes the thermal entropy $S(\beta)=S(\rhov)$ of the ensemble, the first law of thermodynamics holds,
\begin{equation}
 F(\beta) = E(\beta) - \beta^{-1}S(\beta) \ ,
\end{equation}
with the free energy $F(\beta)=-\beta^{-1}\log Z(\beta)$ and thermal energy $E(\beta)=\langle H\rangle_\rhov$. 

We now move to consider comparisons between density matrices $\rhov$ and  $\sigv$. Suppose $\sigv = Z^{-1}(\beta)\exp (-\beta {\bf H})$ and $\rhov = \sigv + \delta \rhov$ 
by a small perturbation $\delta \rhov$ with ${\rm Tr}(\delta \rhov)=0$. We define the change in entropy
\begin{equation}
\delta S \equiv S(\rhov)- S(\sigv) = S(\rhov)-S(\beta)
\end{equation}
and a change in energy
\begin{equation}
\delta E \equiv \langle {\bf H} \rangle_\rhov - \langle {\bf H} \rangle_\sigv = {\rm Tr}(\rhov {\bf H})-E(\beta) \ .
\end{equation}
Then, it is known that the relative entropy (Kullback-Leibler divergence) between $\rhov,\sigv$,
\begin{equation}
K[\boldsymbol{\rho}||\boldsymbol{\sigma}]={\rm Tr}[\boldsymbol{\rho}({\rm ln}\boldsymbol{\rho}-{\rm ln}\boldsymbol{\sigma})]
\end{equation}
for the infinitesimal variation $\rhov = \sigv +\delta \rhov$ from the thermal density matrix $\sigv$ gives an infinitesimal change in free energy $\delta F$ with (compare with \cite{Estrada:2012,Estrada:2012b})
\begin{equation}\label{deltaF}
\beta\delta F =K [\rhov || \sigv] = \delta E - \beta^{-1} \delta S \ .
\end{equation}
Next we proceed to show that by analogous reasoning similar results can be obtained for q-generalised states.

\subsubsection{Thermodynamics corresponding to  q-generalised states with $q>1$}

It is possible to generalise the above derivation based on Gibbs states with $q=1$ for q-generalised states corresponding to the index $q>1$. One starts with 
the constrained extremisation problem, now with the goal of maximising the generalised q-entropy (Tsallis entropy) of form \cite{Tsallis:1988,Abe:2003b,Abe:2004,Abe:2003,Abe:2006,Tsallis:1998}
\begin{equation}
S_{q}[{\bf \boldsymbol{\sigma}}]=\frac{1}{1-q}({\rm Tr}{\boldsymbol{\sigma}}^{q}-1).
\label{qEntropy}
\end{equation}
In the limit $q\rightarrow1$ the q-entropy reduces to the usual von Neumann entropy (\ref{SvN}). Now we add a constraint that the expecation value $\langle {\bf H} \rangle_{\sigv} = {\rm Tr}(\sigv {\bf H})$ is kept fixed and ask what is the form of the density matrix $\sigv$ that maximises entropy in Eq. (\ref{qEntropy}). The result turns out to be $\sigv = \rhov_q$, with \cite{Abe:2003,Abe:2006}
\begin{equation}\label{rhoq}
\rhov_q  =Z_q^{-1}\exp_q ( {-\beta {\bf H}} )\ ,
\end{equation}
where $Z_q = Z_q(\beta) ={\rm Tr}[\exp_q ( {-\beta {\bf H}} )] $. Note that setting ${\bf H}=-{\bf A}$,  Eq. (\ref{rhoq}) agrees with Eq. (\ref{densityMatrix}) as expected.
This gives a q-generalised density matrix in Eq. (\ref{densityMatrix}) for description of the networks with adjacency matrix ${\bf A}$.

Next we will consider two closely related networks, the original modular one with a state $\sigv$ and the rewired counterpart with $\rhov = \sigv + \delta \rhov$. The original 
state is q-distributed, with $\sigv$ of the form in Eq. (\ref{densityMatrix}). We are interested in finding the q-generalised relation of (\ref{deltaF}) between changes in free energy, energy, and entropy. The q-distribution arises from constrained maximisation of the Tsallis q-entropy, generalising the von Neumann entropy. Correspondingly, the appropriate quantity that generalises the Kullback-Leibler divergence is the q-generalised Kullback-Leibler-Tsallis divergence (henceforth
q-divergence for brevity), given by \cite{Abe:2003b,Abe:2004,Abe:2003,Abe:2006} 
\begin{equation}
K_{q}[\boldsymbol{\rho}||\boldsymbol{\sigma}]=\frac{1}{1-q}[1-{\rm Tr}(\boldsymbol{\rho}^{q}\boldsymbol{\sigma}^{1-q})],\label{qDivergence}
\end{equation}
which at the limit $q\rightarrow1$ approaches the Kullback-Leibler divergence. 
The q-divergence is always positive, and zero for identical states.
Note that when $q\in\,]1,2]$ and density matrices are positive semidefinite, the term $\boldsymbol{\sigma}^{1-q}$
needs to be interpreted as $[\boldsymbol{\sigma}^{-1}]^{q-1}$, where $\boldsymbol{\sigma}^{-1}$ is the pseudo-inverse of density matrix
$\boldsymbol{\sigma}$. Here, we have followed the ordering of density matrices $\boldsymbol{\rho} \leq \boldsymbol{\sigma}$ so that states of
$\boldsymbol{\rho}$ (corresponding to the rewired networks) are obtained from $\boldsymbol{\sigma}$ (corresponding to the original modular network), meaning that the number of links in the rewired network is always equal or smaller than in the original one. 

For the q-generalisation of the infinitesimal first law in Eq. (\ref{deltaF}), we follow \cite{Abe:2003}.  We first need a q-generalisation of the expectation value of the Hamiltonian,
defined by \cite{Abe:2003}
\begin{equation}
\bar{E}_{q}={\rm Tr}({\bf H}\boldsymbol{\rho}^{q})/{\rm Tr}(\boldsymbol{\rho}^{q}) \ ,
\end{equation}
the trace in the denominator is needed for an expectation value with respect to a unit normalised density matrix $\rhov^q/{\rm Tr (\rhov^q)}$. With this definition,
by a direct calculation one obtains \cite{Abe:2003} an extension of the usual thermodynamic relation,
\begin{equation}
\delta S_{q}/\delta\bar{E}_{q}=\beta,\;\;\;q\in\,]1,2]\:\:{\rm and}\:\:\beta>0
\label{minusBeta} \ ,
\end{equation}
in agreement with the interpretation of $\beta$ as the inverse temperature.
Starting from the q-divergence as defined in Eq. (\ref{qDivergence}) and assuming that changes $\delta\boldsymbol{\rho}$ in state $\boldsymbol{\rho}$
are small enough so that ${\rm Tr}\delta\boldsymbol{\rho}\approx0$
and $\boldsymbol{\rho}\approx\boldsymbol{\sigma}$ in all cases, we
can show by direct calculation (compare with derivation in ref. \cite{Abe:2003})
that 
\begin{equation}
({\rm Tr}\boldsymbol{\sigma}^{q})\;\delta K_{q}[\boldsymbol{\rho}||\boldsymbol{\sigma}]=\beta\;\delta\bar{E}_{q}-\delta S_{q}[\boldsymbol{\rho}],\:\:\:q\in\,]1,2].\label{Keq1}
\end{equation}

It is now possible to interpret the term $({\rm Tr}\boldsymbol{\sigma^{q}})\;\delta K_{q}[\boldsymbol{\rho}||\boldsymbol{\sigma}]$
as analogous to a change in free energy of a thermodynamic systems, by defining a change in q- free energy as 
\begin{equation}
\delta F_{q}=\beta^{-1}({\rm Tr}\boldsymbol{\sigma}^{q})\;\delta K_{q}[\boldsymbol{\rho}||\boldsymbol{\sigma}].\label{qFreeEnergy1}
\end{equation}
Then, a relation resembling the infinitesimal form of the first law of thermodynamics holds,
\begin{equation}\label{deltaFq}
\delta F_{q}=\delta\bar{E}_{q}-\beta^{-1}\delta S_{q} .
\end{equation}
We now restore the adjacency matrix by replacing ${\bf H}=-{\bf A}$, and denote its q-expectation value by $\bar{A}_{q}={\rm Tr}({\bf A}\boldsymbol{\rho}^{q})/{\rm Tr}(\boldsymbol{\rho}^{q})$, so that $\delta\bar{E}_q=-\delta\bar{A}_q$. The generalised first law then reads
\begin{equation}
\delta F_{q}=-\delta\bar{A}_{q}-\beta^{-1}\delta S_{q} \label{qFreeEnergy2} \ .
\end{equation}
In the alternative interpretation, where the adjacency matrix is chosen as the Hamiltonian, {\bf H'} = {\bf A}, we would have arrived at this form. 
The minus sign in front of $\bar{A}_{q}$ is consistent with the inverse temperature $\beta'=-\beta$ being negative in the alternative interpretation. In that case,
the alternative form of (\ref{deltaFq}) is rewritten as
\begin{equation}\label{deltaFqprime}
\delta F'_{q}=\delta\bar{E'}_{q}-(\beta')^{-1}\delta S_{q},
\end{equation}
where we have defined  $\delta F'_{q}=(\beta')^{-1}({\rm Tr}\boldsymbol{\sigma}^{q})\;\delta K_{q}[\boldsymbol{\rho}||\boldsymbol{\sigma}]=-\delta F_q$ as a change in q- free energy.

At the limit $q\rightarrow1$ the result in Eq. (\ref{qFreeEnergy2}) agrees with the results and interpretation suggested by Estrada \cite{Estrada:2012,Estrada:2009,Estrada:2012b},
who has proposed a similar relation by using the exponential adjacency kernel and Kullback-Leibler divergence. The result in Eqs. (\ref{Keq1})-(\ref{qFreeEnergy2})
can be taken as q-generalisation of the previous results by Estrada.
The advantage of the q-generalised adjacency kernels is that using the freedom provided by choice of parameters $\beta$ and $q$, the
contribution of paths of different lengths to the connectivity (or communicability) of nodes can be explored. At the limit $q\rightarrow1$
short paths are more dominant for connectivity than long ones, while at the limit $q\rightarrow2$ the contribution of longer paths becomes
more important. In both cases, also weight factors can be used to tune the scale of paths that one wishes to take into account in the
connectivity of nodes. At the limit $\beta\rightarrow0$, a vanishing weight is attached to all links, so that the system is essentially
a set of disconnected nodes, while the larger the value of $\beta$, more tightly connected the system.

\subsection{Generalised free energy and response function from simulations}

The viability of the thermodynamic-like description of heavy-tailed networks is tested by first generating a set of networks using the generative
model, designed to produce networks with inverse power law type degree distribution with power $1<\lambda<2$ and with high modularity. To
obtain the q-generalised free energy from simulations is, however, not entirely straightforward for two reasons. First, the systems are
always finite and thus the density matrix as defined in Eq. (\ref{densityMatrix}) does not exactly maximise the q-entropy. Secondly, due to the small
size of the systems, statistical fluctuations are large, and it is difficult to obtain the quantities ${\rm Tr}\boldsymbol{\sigma}^{q}$,
$\delta S_{q}$ and $\delta\bar{A}_{q}$ accurately, while change in divergence $\delta K_{q}$ is computationally more stable. Therefore,
based on the simulations where networks are generated, we first justify that relation in Eq. (\ref{Keq1}) holds by demonstrating the validity
of equivalence 
\begin{equation}
K_{0}(\beta)\;\delta\tilde{K}_{q}=-\beta A_{o}(\beta)\delta\bar{A}_{q}-\delta S_{q}.\label{eq3}
\end{equation}
Here we have assumed that $K_{0}\simeq{\rm Tr}\boldsymbol{\sigma}^{q}{\rm Max}[\delta K_{q}]$ depends only on $\beta$, and $\delta\tilde{K}_{q}=\tilde{K}_{q}/{\rm Max}[\tilde{K}_{q}]$
only on rewiring frequency $\pi$. The normalisation parameter $A_{0}(\beta)$ is allowed to compensate for the $\beta$-dependence of the normalisation
factor ${\rm Tr}\boldsymbol{\rho}^{q}$, due to the finite size of networks affecting the cut-off of eigenvalues of the adjacency matrix.
The validity of Eq. (\ref{Keq1}) is then guaranteed in the region of parameters where the scaling functions $K_{0}$ and $\tilde{K}_{q}$
can be found so that Eq. (\ref{eq3}) holds. As will be seen, this is possible for networks with heavy tails characterised by $1<\lambda<2$
when $1.3<q<1.7$ (in some cases also for lower and higher values of $q$). In this region of parameter $q$, the thermodynamic-like description
is viable and the q-generalised free energy change is 
\begin{equation}
\delta F_{q}=\beta^{-1}K_{0}\;\tilde{K}_{q},\label{qFreeEnergy3}
\end{equation}
where functions $K_{0}$ and $\tilde{K}_{q}$ are obtained by fitting to simulation data.

The q-generalised free energy allows us to define how the network responds to changes in the external parameters $\beta$ and $\pi$.
In the thermodynamic interpretation, the response to changes in $\beta$ is analogous to heat capacity 
\begin{equation}
\chi=\beta^{2}\frac{\partial^{2}(\beta F_{q})}{\partial\beta^{2}}\propto\beta^{2}\,\frac{\partial^{2}K_{0}}{\partial\beta^{2}},
\end{equation}
where it is assumed that the scaling with regard to $\beta$ is the same for the rewired and non-rewired (fully modular) networks. Then, the
response is determined up to an unknown constant (only the change in free energy due to rewiring is known). The response function allows
a simple description how and in what state the network is most responsive to changes in the strength of the links. Different definitions of
the response function are possible (see e.g. \cite{Fronczak:2006}), but here we have preferred to retain the analogy with a thermal response
(i.e. heat capacity).

\section{Results and discussion}

The results for the diagonal values of the q-adjacency kernels are discussed first, because they have the closest relation to a thermodynamic-like interpretation in the case  $q =1$, when diagonal values provide the so-called Estrada index, which yields to a thermodynamic interpretation in terms of free energy \cite{Estrada:2012,Estrada:2012b}. In small systems, however, the distribution of eigenvalues has so large fluctuations that the utility of such interpretation is in practice diminished. Here, we show results that demonstrate that a thermodynamic-like interpretation similar to that proposed by Estrada for diagonal components is viable also for generalised q-adjacency kernels, which describe the state of the network as a whole, but with much reduced fluctuations. 

\subsection{Choice of parameters}
The simulation results to be discussed here show that within a certain
region of parameters $q$ and $\beta$ we can maintain the interpretation
of divergence difference $\Delta K_{q}$ as change of free energy
$\Delta F_{q}$ of the network (note that for simulation results,
symbols $\Delta(\cdot)$ instead of $\delta(\cdot)$ without subscript
$q$ are used). 
One of the central choices is the choice of $\beta_{{\rm MAX}}$,
which must be lower than the inverse of the highest eigenvalue of
the adjacency matrix. Since the highest eigenvalue of the adjacency
matrix depends on the details of the network, the exact choice of
maximum value is case-dependent. In practice, however, for the heavy-tailed
networks of interest here, values of $\beta$ which are from 50\%
to 70\% of the maximum value $\beta_{{\rm MAX}}$ provide the best
compromise between stability of results and resolving power \cite{Lommi:2019}.
On the other hand, when $\beta$ is too low, resolving power with
regard to differences in characteristics of nodes is lost. Therefore,
the $\beta_{{\rm MAX}}$ is first chosen to be about 70\% of the highest
possible value, and it depends on $\lambda$ linearly so that $\beta_{{\rm MAX}}=-0.02+0.33\lambda$
for $\lambda\in]1,2]$. The values of $\beta_{{\rm MAX}}$ for different
choices of $\lambda$ are summarised in Table I. The adjacency kernel
for each value of $\lambda$ is then evaluated for seven values ranging
from $\beta_{{\rm MAX}}$ to 0.7$\beta_{{\rm MAX}}$.

\vspace{0.25cm}

\begin{table}[h!]
\setlength{\tabcolsep}{3pt} \centering \caption{The parameters $\beta_{{\rm MAX}}$ in simulations and the maximum
number of attempted links $M_{{\rm A}}$ and realised links $M_{{\rm R}}$ (on
average) with $M_{0}$=2000. In all cases, the number of connected
nodes is on average about N=1000. The realised powers $\lambda'$
for given $\lambda$ are also shown. }
\label{table1} \bigskip{}
\centering 
\begin{tabular}{lccccc}
\hline 
\noalign{\smallskip{}
}  & \hspace{0.5cm}  & $\lambda=1.3$  & $\lambda=1.5$  & $\lambda=1.7$  & $\lambda=1.9$ \tabularnewline
\noalign{\smallskip{}
}\noalign{\smallskip{}
} $\beta_{{\rm MAX}}$  &  & 0.023  & 0.030  & 0.036  & 0.043 \tabularnewline
${\rm M}_{{\rm A}}/{\rm M}_{0}$  &  & 2.0  & 1.4  & 1.1  & 1.0 \tabularnewline
${\rm M}_{{\rm R}}/{\rm M}_{0}$  &  & 1.8  & 1.3  & 1.0  & 0.9 \tabularnewline
$\lambda'$  &  & 1.7  & 1.8  & 1.9  & 2.1 \tabularnewline
\hline 
\end{tabular}
\end{table}

\vspace{0.25cm}

In the simulations to produce generic networks, only one type of modularity
is chosen, corresponding to the three-tiered modular structure where
the first tier has units of N'=200 nodes, the second tier three
of such units of 200 nodes (for a total of 600), and the third tier three units
of 600 nodes, providing N=1800 nodes in total. All nodes, however, are not connected and the choice of parameters lead to total
number of connected nodes of about N=1000. The structure
is thus a block model with a three-tiered hierarchy of connections.
This selection results in the initial modular structure with modularity
in range $0.80<Q<0.90$ depending on the case. As was explained in section 2, the parameters are selected so that they roughly correspond to
values found empirically for networks of associative, thematic knowledge \cite{Koponen:2018b,Lommi:2019}.
However, provided that the modularity is high enough, the exact structure of initial modularity and its variation are not crucial for the properties of interest
here (for effects of modularity, see ref. \cite{Koponen:2018c}). Therefore one initial state is enough. Instead, the relaxation of
the initial modularity and its effects are of interest. The modularity of the networks is relaxed by rewiring the links with relative frequency $\nu \in [0,\nu_{{\rm MAX}}/M]$. In practise, the value
$\nu_{{\rm MAX}}=5.0\cdot10^{5}$ is chosen to guarantee complete rewiring, while $\nu=1$ means that, on average, each link is rewired once. 
In the rewiring, performed with \texttt{IGraph} algorithm \texttt{IGRewire}, the degree sequence of the nodes is preserved.

\subsection{The diagonal values of density matrices}

The degree centrality distributions $P(d)$ for degrees of nodes are shown in Figure 1 (the upper row). The distributions are clearly heavy-tailed
and can follow a power law in a broad range of degrees. It should be noted that irrespective of the choice of $\lambda$, defining the
affinity distribution, the powers corresponding to the degree distribution are rather narrowly distributed from $\lambda'\approx1.7$ for $\lambda=1.3$
up to $\lambda'\approx2.0$ for $\lambda=1.9$. This behaviour is most likely caused by the fact that due to constraints set by modularity,
high affinity nodes are more likely to be connected when only about 1000 of all the potential 1800 nodes are connected. A similar effect
of values $\lambda'$ exceeding the values of $\lambda$ was also observed in simulations where a broader set of modularities was explored
\cite{Koponen:2018c}.

The density distribution of diagonal values $[\boldsymbol{\rho}]_{ii}$ of density matrix $\boldsymbol{\rho}$ are related to the eigenvalues
of the density matrix and are shown in two lower rows in Figure 1
for the fully modular and fully rewired networks. The diagonal values
have different distributions corresponding to different values of
$q$ and thus show that they are more sensitive to the details of
the networks than the degree distribution. Moreover, the relaxation
of modularity substantially affects the leading edge of the distribution.
The modular networks have high number of low values of $[\boldsymbol{\rho}]_{ii}$
due to modular structure, in which local connections in modules are
tight. However, with relaxation of the modularity, the distributions
become more power-law type and higher values of $[\boldsymbol{\rho}]_{ii}$
become more dominant, indicating that the number of longer paths increases
with loss of modularity. 
This effect is more pronounced the higher the value of $q$. For the
lowest values of $q$ the distribution of the diagonal values of $[\boldsymbol{\rho}]_{ii}$
is in all cases close to an inverse power law distribution. For higher
values of $q$, the high frequency leading edge corresponding to the
low values of $[\boldsymbol{\rho}]_{ii}$ begins to deviate from power
law behaviour. Also, in the low frequency part, the statistical fluctuations
are high. In practice, for example in case of associative knowledge
networks, just this part of the distribution is of interest. With
small sample sizes, however, the large statistical fluctuations become
an obstacle for similarity comparisons based on Kullback-Leibler divergence;
the statistical fluctuations are usually too large to allow robust
results \cite{Lommi:2019,Koponen:2018c}.

\begin{figure}
\centering 
\includegraphics[width=14.0cm]{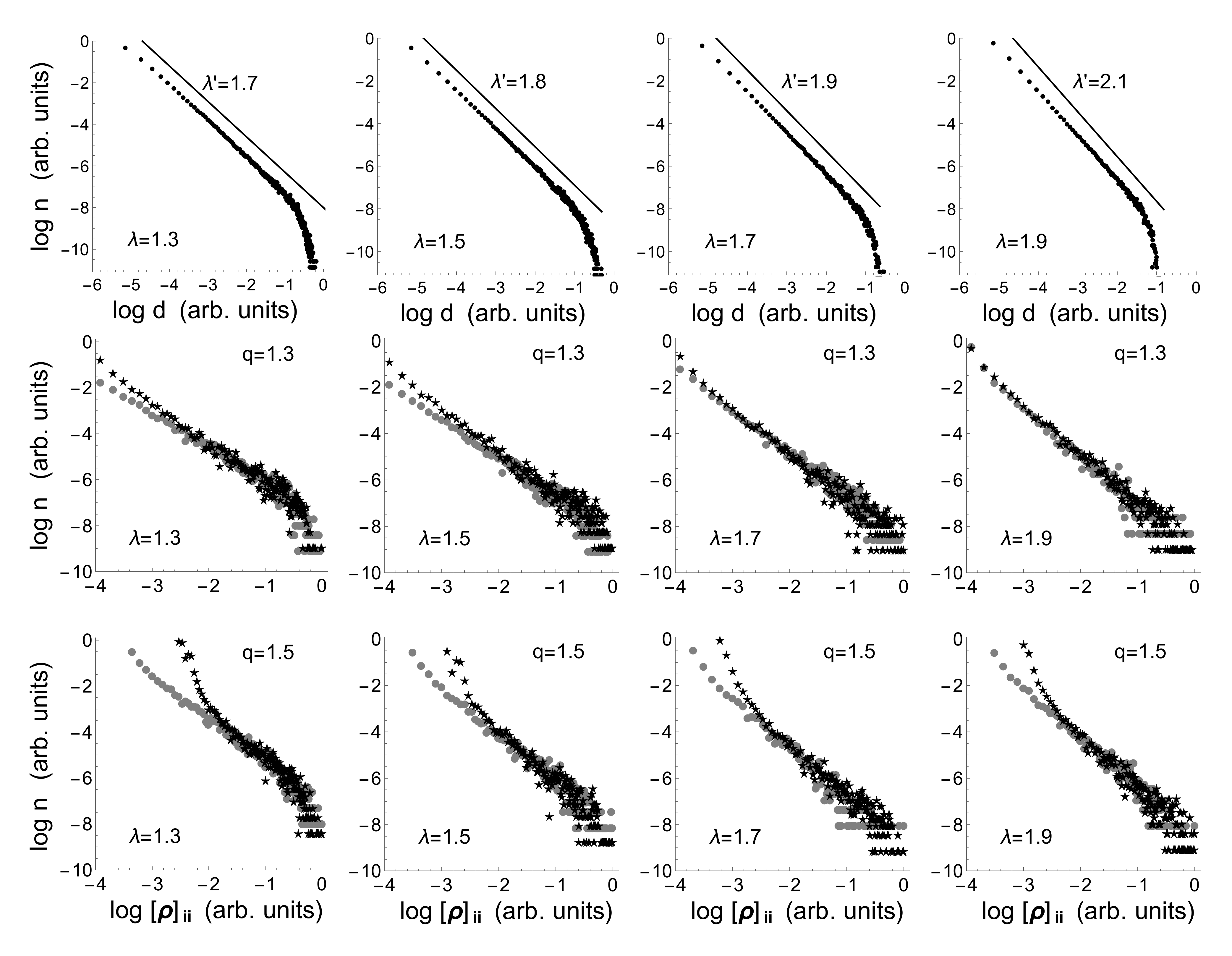}
\caption{Distributions of degree $d$ for parameters $\lambda=$ 1.3, 1.5,
1.7 and 1.9 (upper row, with estimates of power-law fits shown by
$\lambda'$). The distributions of diagonal elements $[\boldsymbol{\rho}]_{ii}$
of density matrix $\boldsymbol{\rho}$ for each $\lambda$ and values
$q=$ 1.3, 1.5 and 1.7 are shown on two lowest rows, respectively.
In each case, the distributions are shown for original modular (symbol
$\star$, upper sets of data points) and completely rewired networks
(symbol $\bullet$, lower sets of data points) with $\beta=\beta_{{\rm max}}$.
The results for degree distributions are with $10^{4}$ repetitions
while for $[\boldsymbol{\rho}]_{ii}$ 100 repetitions are used.}
\label{fig1} \vspace{0.5cm}
\end{figure}

\subsection{The thermodynamic-like interpretation}

\begin{figure}
\centering 
\includegraphics[width=14.5cm]{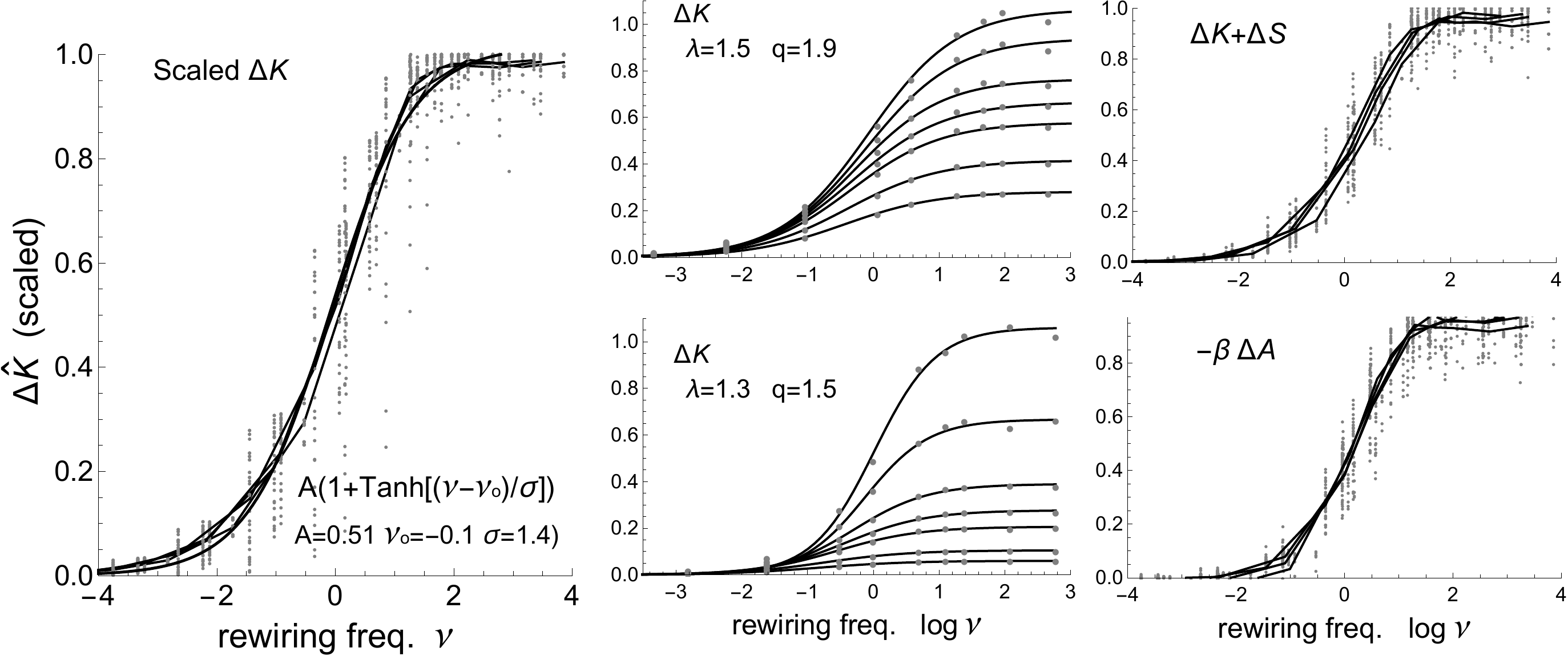}
\caption{The scaled divergence difference $\Delta \hat{K}$ (see text for definition)
is shown on the left panel as a function of rewiring frequency $\nu$.
In the middle panel are shown values of $\Delta K$  (scaled to maximal value in the ensemble)
for heavy-tailed networks with different values of $\lambda$. At right are shown the (scaled) values of $\Delta K + \Delta S$ (the upper figure) and $-\beta \Delta A$ (the lower figure).  
Note that interpretation of $\Delta K$
as free energy requires that these terms are linearly proportional.
In all cases, data-points are plotted showing the variance in data.
The natural logarithm ${\rm Log}$ is used throughout.}
\label{fig2} \vspace{0.5cm}
\end{figure}

To compare heavy-tailed networks and the effects of modularity we
use the q-generalised entropy, q-generalised divergences, and the
complete density matrix $\boldsymbol{\rho}$ to characterise the networks.
In addition, a suitably chosen q-adjacency kernel may allow macroscopic
description based on q-generalised free energy. To secure this, we
need to show that the relation in Eq. (\ref{eq3}) holds for simulation
results. Figure 2 shows the change $\Delta K_q$ (simulation results
for $\delta K_{q}$ and with subscript $q$ omitted in what follows) for different values of
$q$ and $\beta$ as a function of rewiring frequency $\nu$ so that
all divergences are scaled to the maximal value $K_{0}={\rm Max}[\Delta K]$. 
The reference (density matrix $\boldsymbol{\sigma}$ in Eq. (\ref{qDivergence})) is chosen
to be the original, fully modular network. The curves for $\Delta K$
are sigmoidal, but their maximum values $K_{0}={\rm Max}[\Delta K]$
are different for different choices of $q$ and $\beta$ as shown
in Fig. \ref{fig2} in the middle panel. However, the scaled values $\Delta \hat{K} =\Delta K/K_{0}$
collapse to a single curve (shown at left in Fig. \ref{fig2}) that is the same for all
choices of $q\in[1.3,1.9]$ and $\beta/\beta_{{\rm MAX}}\in[0.7,1.0]$,
thus demonstrating the scaling of the $\Delta K$. The transitional
region where the properties of the network change essentially is now
recognised to be located in $-1<{\rm Log} \, \nu<1$ and in region ${\rm Log}\,  \nu>2$
(note that here ${\rm Log}$ refers to natural logarithm) the network
is essentially completely rewired (i.e. random, only degree sequence
preserved). The most interesting properties of the changes of the
network, however, are contained to the maximal value $K_{0}(\beta;q,\lambda)={\rm Max}[\Delta K]$,
which depends on $\beta$ and parameters $q$ and $\lambda$. This
quantity is discussed in more detail below.

To demonstrate the validity of Eq. (\ref{eq3}) the scaled values of
simulation results for $\Delta K+\Delta S$ and $-\beta\Delta A$
(corresponding to quantities at left and right side in Eq. (\ref{eq3}),
respectively) are shown separately on the panel at right of Fig. \ref{fig2}.
If Eq. (\ref{eq3}) holds, the scaled curves should be linearly proportional.
This is suggested by the appearance of the curves shown. The curves,
however, are averaged for each set of parameters over the 100 repetitions
where the statistical fluctuations are still large, as shown in Fig.
2. The Pearson correlation of data points for $\Delta K +\Delta S$
and $-\beta \Delta A$ is in all cases from 0.97 to 0.99 with p-values
well below $10^{-4}$, indicating good correlation in most cases,
and at worse close to $10^{-3}$ for high values of $q$, still indicating
a reasonable correlation. This means that we can take the relation
in Eq. (\ref{eq3}) to be valid in the cases studied. The dependence
is also linear with good accuracy; the residuals of the averaged values
from linear fits remain below 10\% of the mean values for the average
values shown in Fig. \ref{fig2} for $q<1.5$, increasing to about
30\% in the worst cases for $q=1.9,\lambda=1.3$ and $\beta\approx\beta_{{\rm MAX}}$.
The results indicate that the interpretation of the divergence in
terms of free energy as suggested by Eqs. (\ref{qFreeEnergy2}) and
(\ref{qFreeEnergy3}) is viable provided that parameter $q$ is in range
$1.3<q<1.7$ and close enough $\lambda$ for $q<1.7$. Values $q<1.3$
could not be tested, because computations rapidly become unstable
for low values of $q$.

\begin{figure}
\centering 
\includegraphics[width=14.0cm]{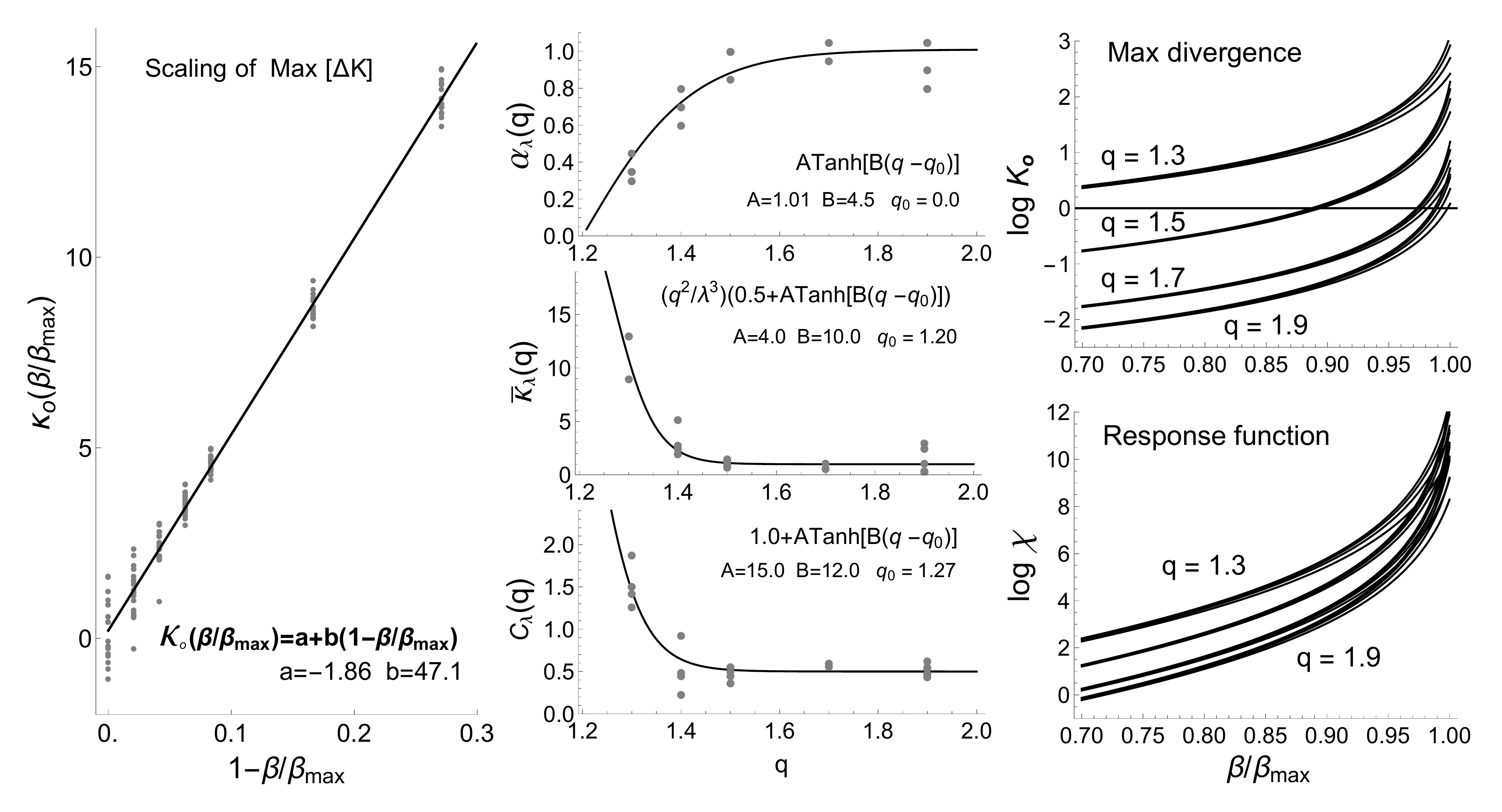}
\caption{The scaling functions for construction of free energy $F_q$. 
The dependence of maximum value $\kappa_{0}$ of $\Delta K$ is shown
at the left panel, in scaled form and as function of $1-\beta/\beta_{{\rm max}}$. 
On the middle panel we show the fits to empirical data as used to construct
$\alpha$, $\bar{\kappa}$ and $C_{o}$ for the semi-empirical functional
form of free energy $F_q$. In these plots, the data-points are shown.
On the right panel we show the semi-empirical construction of functional
form of $\kappa_{0}$ for the maximum value of change in divergence
(upper figure) and the corresponding function for the thermal response
$\chi$. In both figures, the dominant dependence is on parameter
$q$ and within each set of results, corresponding to given choice
of $q$, a bunch of results corresponding to different choices of
$\lambda$ is visible. In each bunch, the highest value of $\lambda=1.9$
corresponds to the topmost curve and in the curves below it correspond
to the lower values $\lambda$ = 1.7, 1.5 and 1.3 in that order.}
\label{fig3} \vspace{0.5cm}
\end{figure}

The dependence of the maximal value $K_{0}={\rm Max}[\Delta K]$
on $\beta$ and parameters $q$ and $\lambda$ is somewhat complex
but regular. In Figure \ref{fig3}, at the left panel, $K_{0}$ is
shown scaled in a suitable way, to obtain a linear dependence. The
linear form $\kappa_{0}(\beta/\beta_{{\rm MAX}})=a+b(1-\beta/\beta_{{\rm MAX}})$
with $a=-1.86$ and $b=47.1$ shown in Fig. \ref{fig3} is obtained
by scaling 
\begin{equation}
\kappa_{0}=\bar{\kappa}_{\lambda}K_{0}^{-\alpha_{\lambda}}-C_{\lambda}.
\end{equation}
Here coefficients $\bar{\kappa}_{\lambda}(q)$ and $C_{\lambda}(q)$
and power $\alpha_{\lambda}(q)$ depend on parameters $\lambda$ and
$q$ as shown in Fig. \ref{fig3} in the middle panel. The sigmoidal
fitting functions displayed in the figures for these parameters are
obtained from simulation results through simple, descriptive fits
to data. It should be noted that there is no theoretical basis to
motivate the form of the fits so they are only practical vehicles for
interpolation and moderate extrapolations. Moreover, the number of
parameter combinations feasible and reasonable here is quite limited,
which means large insecurity in the fits. However, these limitations
are not severe, because we are not aiming here at quantitatively accurate
description but rather to demonstrate the generic behaviour of the
networks when control parameters are changed. The generic behaviour
is adequately obtained despite the limited accuracy of the fits. The
linear behaviour of $\kappa_{0}$ with the fitting function for the
corresponding coefficients allows us to find the functional dependence
of $\Delta K$ on $\beta$, needed for macroscopic description of
the network.

The scaling function $\kappa_{0}$ allows us finally to obtain the
maximum value $K_{0}$ and through it, the response function $\chi$,
which for a true thermodynamic system would correspond to the thermal
capacity. The maximal value $K_{0}$ and response function $\chi$
are shown in Figure \ref{fig3} at right. Their most important feature
is that they increase with increasing value of $\beta$, which means
that with higher $\beta$ better resolving power (larger differences)
is obtained. Also, the response function allows the interpretation
that the larger the $\beta$, the more sensitive the value of divergence
is to changes in structure of the network (i.e. with a large value
of $\beta$ better resolving power is possible). Another interesting
notion is that the value of $q$ mostly determines the behaviour of
both, while values of $\lambda$ affect them much less; the results
are bunched according to the value of $q$, and within these bunches,
small differences originate from different choices for parameter $q$.
Roughly, this means that differences in $\Delta K$ are more pronounced
the smaller the value of $q$. This, of course, is expected in the
case of heavy-tailed distributions since the smaller the values of
$q$, the more sensitive the divergence is for low frequency values
in distributions. Note, however, that if one wishes to retain the
validity of Eq. (\ref{eq3}) (and thus, also the validity of Eqs. (\ref{qFreeEnergy2})
and (\ref{qFreeEnergy3})) the value of $q$ for divergence cannot be
chosen freely but must be same as for the q-adjacency kernel.

The present study leaves unanswered the question how the current model behaves when $q\rightarrow1$. The theoretical
argumentation presented here suggests that at that limit one should arrive at a description based on canonical ensemble (Gibbs' ensemble),
von Neumann entropy and Kullback-Leibler divergence. However, 
for values  $\lambda<1.3$ reaching good computational stability is challenging and computations become quite unstable. Similarly, when of $q \rightarrow 2$ the computational results become
very unstable. In both cases, reaching good stability would require very extensive ensembles. The detailed reason for this is not
known to us, but such behaviour is compatible with the notion that extrapolation from non-extensive thermodynamics of finite systems
($q>1$) to the usual canonical extensive thermodynamics ($q=1$) also requires the thermodynamic limit of
infinite systems (compare with ref. \cite{Abe:2003}).

In summary, the main result contained in Figs. \ref{fig2}-\ref{fig3}  is that the theoretically defined q-generalised free energy is a viable description of the macro-level state of the network
changes when the modularity of the network is reduced. Within the thermodynamic-like interpretation, the change can be interpreted as
lowering of the free energy due to relaxation of the modularity. In connection with this change, the probability of long paths increases.
The effect is best seen for the adjacency kernels with low values of $q<1.7$. The model also contains parameter $\beta$, which allows
a different weighting of links, $\beta>1$ corresponding to strong links, while $\beta<1$ means making the links weaker, with the extreme
case $\beta\rightarrow0$ corresponding to totally disintegrating the network.

\section{Conclusions}

The practical motivation to conceptualise the properties of heavy-tailed
networks from the macroscopic viewpoint derives from the notion that
associative knowledge networks tend to be heavy-tailed, with broad
tails that can be fitted with inverse power laws with powers in the
range $1<\lambda<2$ and that in addition these networks are often
highly modular \cite{Koponen:2018b,Lommi:2019,Morais:2013}. However,
real empirical networks are often quite small, and they cannot be
easily or reliably characterised by pre-selected distributions of
certain centrality like communicability or Katz centrality. Also,
the similarity comparisons are often very awkward \cite{Koponen:2018b,Lommi:2019}.
This prompts an attempt to use more holistic descriptions based on
the network's density matrix, which provides all available information
about the network \cite{Biamonte:2019,Domenico:2016}. 
Such description
can be based on q-generalised adjacency kernels introduced here, which allows interpolation between known 
exponential and Neumann kernels. 

The q-generalised  adjacency kernel opens a door to the macro-level description of heavy-tailed networks, through 
the q-generalised Kullback-Leibler-Tsallis divergence.  A parallel result has previously been obtained for closely similar
q-generalised ensembles in the case on non-extensive statistics in general  \cite{Abe:2003,Abe:2006}. Here we have shown that analogously, the Kullback-Leibler-Tsallis divergence can be interpreted 
as a change of a q-generalised free energy of a modular heavy-tailed complex network, when modularity becomes relaxed. 
The finding that it is possible to define q-generalised free energy of heavy-tailed networks, and based on it, to obtain a thermodynamic-like interpretation, opens up interesting avenues
to describe networks using thermodynamic-like concepts. For example, it is shown that  with the q-generalised free energy it is possible to
derive the response of the network for changes of values of $\beta$, paralleling the thermal response if the thermodynamic interpretation
is evoked. The response of the network with increasing value of $\beta$ indicates that changes in the rigidity of the network (resilience)
are smaller, the larger the values of $\beta$; the thermal capacity of the network increases when $\beta$ increases.

The present study leaves many fundamental questions unanswered, most notably the question how the role of the adjacency matrix should be interpreted in terms of the Hamiltonian, and how far one can pursue 
the physical interpretations on the basis of structural, mathematical analogies. Nevertheless, we believe that the results provided here are promising steps towards a more general theory of heavy-tailed networks.

\bigskip{}
\bigskip{}

\clearpage{ }

\section*{Appendix A: The q-communicabilities and q-divergence}
\setcounter{equation}{0}
\setcounter{figure}{0}
\renewcommand\theequation{A.\arabic{equation}}
\renewcommand\thefigure{A.\arabic{figure}}

This Appendix is a sequel to the main text and in it, we discuss  communicability distributions. Starting from the q-generalized kernels, we define the q-generalized
communicabilities. We then study their distributions arising from networks with heavy-tailed degree distributions, and use the q-divergences to differentiate between q-communicability distributions
arising from modular and rewired networks.

The q-exponential adjacency kernel in Eq. \ref{Qkernel} provides a basis for defining a q-generalised communicability (in brief, q-communicability) between nodes $k$ and $i$ as a row sum of q-adjacency kernel, in form

\begin{equation}
\Gamma_{i}(q,\beta) = Z^{-1} \sum_{j}  {\rm exp}_{q} [ \beta \,{\bf A}]_{ij}  \,
\end{equation}
where $Z= {\rm Tr} \;  {\rm exp}_{q}[\beta {\bf A}]  $ is the normalisation factor.  The q-generalised total communicability provides Estrada's total communicability in the limit $q \rightarrow 1$ and the Katz centrality is obtained in the limit $q \rightarrow 2$, corresponding to the exponential and Neumannn kernels, respectively.

For heavy-tailed networks, the q-communicability values are distributed according to an inverse power law, with inverse powers $\gamma$ which are roughly in range from 2 to 3, but with a cut-off corresponding to largest observed value $\Gamma_{\rm max} ={\rm Max}[\{ \Gamma_{i}\}]$.  By rescaling the values of $\Gamma_i$ by the maximum $\Gamma_{\rm max}$, and relabel $\Gamma_i /\Gamma_{\rm max}\rightarrow \Gamma_i \in [0,1]$, the new values can be represented in the form of the discrete distribution
\begin{equation}
\label{Qprob}
p_{i}(\Gamma_i) = N^{-1} \left(1+(\alpha-1)\frac{1}{\epsilon}  \, \Gamma_i \right)^{-1/(\alpha-1)}, \:\: {\rm with} \: \alpha \in \left]1,2\right] ,
\end{equation}
where $p_i$ is the probability of a given value $\Gamma_{i} \in [0,1]$ and  $N$ is the normalisation factor. The exponent $\alpha$ is related to inverse power $\gamma$ as $\alpha=1+1/\gamma$. 
In what follows,  we use $\alpha$ in discussing the theoretical results while for simulation data $\gamma$ is often more convenient.   The parameter $\epsilon=\gamma/ \Gamma_{\rm max} \ll 1$ has small values, but must always be different from zero. The distribution in Eq. \eqref{Qprob} is q-exponential with exponent $q =\alpha$, where $\alpha$ is now used to refer  to the q-index to avoid any confusion with index $q$ in Eq. \eqref{Qkernel} associated with a q-adjacency kernel. 

\begin{figure}
\centering 
\includegraphics[width=14.0cm]{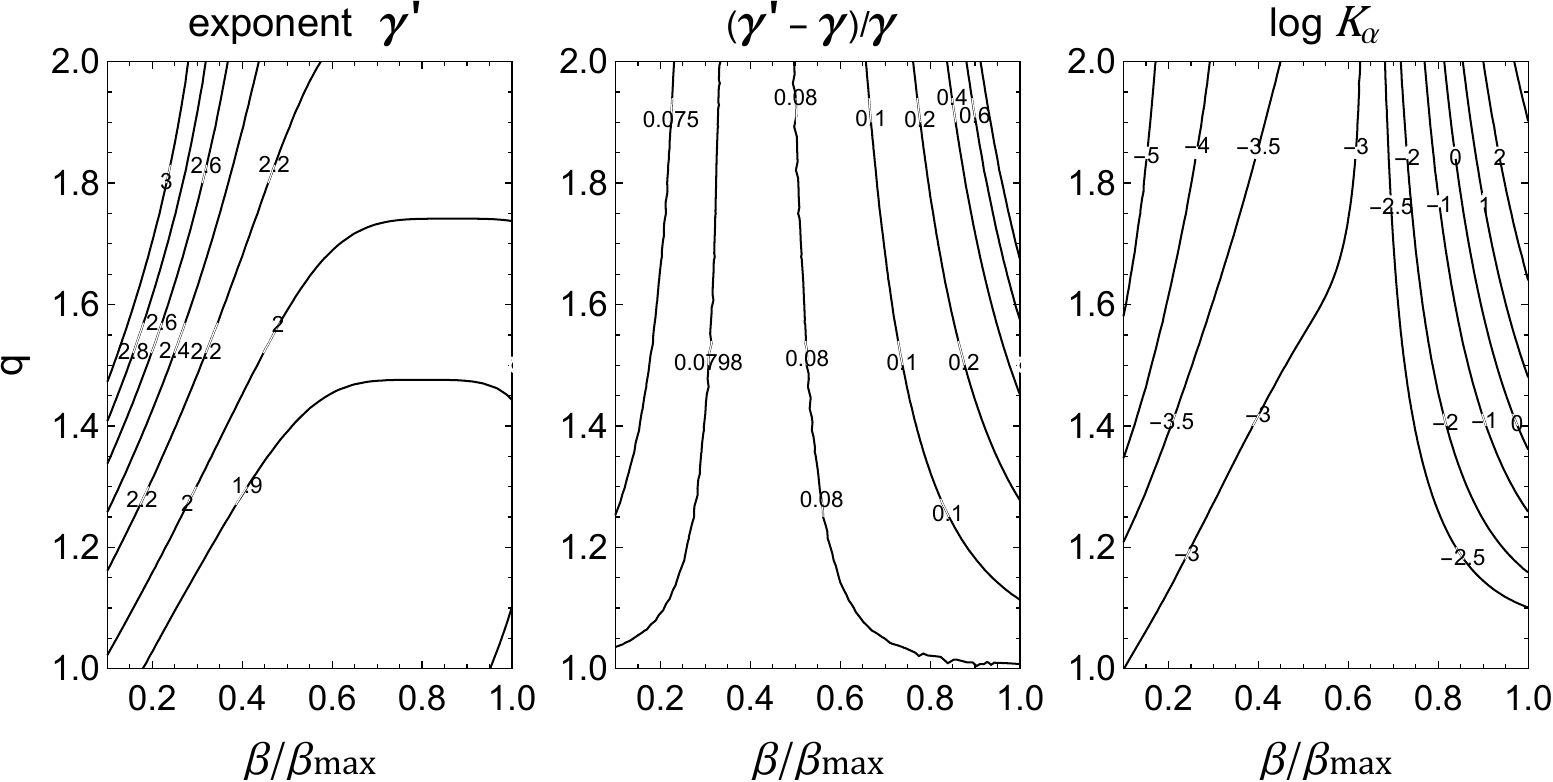}
\caption{The contour plots of inverse powers $\gamma'$ for modular network (at the left) and the relative difference $(\gamma' -\gamma)/\gamma$ (in the middle)  with $\gamma$ corresponding to a non-modular (rewired) distribution. The logarithm of q-divergence $K_\alpha$ is shown at the right. Results are shown for parameters $q \in ]1,2]$ and $\beta/\beta_{\rm max} \in [0.1,1.0]$. Note that the valus of $\alpha$ and $\gamma$ are related through $\alpha=1+1/\gamma$. In reporting inverse powers we have used values of $\gamma$ because the range of their variation is [2,3] while for $\alpha$ the corresponding range is narrower [1.33,1.5]. }
\label{figcontours2} \vspace{0.5cm}
\end{figure}

Results for power $\gamma$ characterising the  q-communicabilities in case $\Lambda=1.5$ are shown in Figure \ref{figcontours2}. The results are obtained for 
$q = \{ 1.05,1.1, 1.2, \ldots, 2.0 \}$ and with $\beta/\beta_{\rm max} = \{0.1, 0.2,0.3,\ldots,1.0\}$, where $\beta_{\rm max}$ is about 80 \% of the inverse of the largest estimated eigenvalue of the adjacency matrix. 
The values of  $\gamma$, and $\gamma'$ in Fig.  \ref{figcontours2} for fully rewired ($Q$=0) and original modular ($Q$=0.86) networks, respectively, are obtained by least-square fits to log-log distributions. In fitting the coefficients, attention is paid to the tail of the distribution, and small values  $\Gamma < 10^{-3}$ are allowed to deviate from the fits. 
 The difference between powers $\gamma$ and  $\gamma'$  is generally quite small, but detectable, as is seen in Fig. \ref{figcontours2} from the relative change $(\gamma'-\gamma)/\gamma$ (in the middle).

We can now utilise the q-divergence to compare q-communicability distributions with better resolution than provided by simply comparing the powers $\gamma$ and $\gamma'$.
With the parametrisation $\alpha= \alpha'(1+\eta)$, where $\eta=(\alpha-\alpha')/\alpha' > 0$  is the relative difference, we can now obtain the q-divergence from Eq. \eqref{qDivergence} by replacing the density matrices with the discrete distributions and the traces of matrices by sums over the distributions, in the conventional way. A closed analytical form for the q-divergence is then obtained by replacing the discrete distribution \eqref{Qprob} with the corresponding continuous distribution and evaluating the sum  in Eq. \eqref{qDivergence} as a Riemann integral. These replacements are not trivial but can, however, be justified for q-exponentials \cite{Plastino:2017}. The resulting expression for the q-divergence is then

\begin{equation}\label{qanal1}
    K_\alpha(\eta,\epsilon) = 
    \frac{1}{\alpha-1}\left[\kappa_\alpha(\eta,\epsilon)\bar{K}_\alpha(\eta,\epsilon)\left(
    \epsilon^{\frac{-(2-\alpha)}{\alpha-1}} - (1+\epsilon)^{\frac{-(2-\alpha)}{\alpha-1}}
    \right)^{\alpha-1} - 1\right]
\end{equation}

\begin{equation}\label{qanal2}
    \kappa_\alpha(\eta,\epsilon) = \frac{(\alpha-1)^{\alpha-1}(2-\alpha(1+\eta))^{\alpha}}{(2-\alpha)^{\alpha-1}[2-\alpha(1+2\eta)][\alpha(1+\eta) - 1]^{\alpha-1}}
\end{equation}

\begin{equation}\label{qanal3}
    \bar{K}_\alpha(\eta,\epsilon) = \frac{\epsilon^{\frac{\alpha(1+2\eta)-2}{\alpha(1+\eta)-1}} - (1+\epsilon)^{\frac{\alpha(1+2\eta)-2}{\alpha(1+\eta)-1}}}{\left(\epsilon^{1-\frac{1}{\alpha(1+\eta)-1}} - (1+\epsilon)^{1-\frac{1}{\alpha(1+\eta)-1}}\right)^{\alpha}}
\end{equation}

The parameter $\eta = (\alpha-\alpha')/\alpha' > 0 $ is the relative difference of the q-indices, characterising the distribution, while the parameter $\epsilon$ is assumed to be the same for both distributions. Note that the parameter $\eta$ attains always small values in the region of interest, where $\alpha=1+1/\gamma$ with $\gamma \in ]2,3]$. 

By using the above results with the values $\gamma$ and $\gamma'$ from Fig. \ref{figcontours2} we eventually obtain the q-divergence $K_{\alpha}$ shown in Fig. \ref{figcontours2} (at the right) as a function of the parameters $q$ and $\beta$. We   also used the analytical results in Eqs. \eqref{qanal1}--\eqref{qanal3} and in them, conversion $\eta = (\alpha'-\alpha)/\alpha$, where  $\alpha=1+1/ \gamma$ and $\alpha'=1+ \gamma'$. The main message of the result in Fig. \ref{figcontours2} is the following: From the values of $K_\alpha$ in Fig. \ref{figcontours2}, we see that although the region close to the Katz centrality (the upper right corner of contour plot with $\beta >0.8$ and $q>1.7$) displays better resolution (i.e. large values of q-divergence), the region corresponding to Estrada's communicability at $q<1.2$ and $\beta >0.3$ also performs reasonably well in providing large enough divergences for good resolution. In this region the choice of $\beta$ is not so crucial for the stability of the calculations as in the case of the region corresponding to the Katz centrality.   

Finally, it should be mentioned that although the thermodynamic-like interpretation of q-divergence in Eqs. \eqref{qanal1}--\eqref{qanal3} is tempting, this is not viable now. The major complication in the thermodynamic interpretation arises now from the fact that communicability is a derived quantity which depends on parameter choices and does not refer to parameter independent and constitutive property of a network. Therefore, the Eq. \eqref{qanal3} is better interpreted as divergence only.  The utility of the analytical result in Eq. \eqref{qanal3} is, however, that it can be used to estimate the optimal region for the parameters with a compromise between a good resolution of divergence and computational simplicity. 

The results demonstrate the practical utility and robustness of Estrada's communicability for applications where heavy-tailed distributions are involved. The results support Estrada's communicability as the most obvious choice, in many cases when a compromise between computational stability, ease, and sufficient accuracy is needed. This conclusion is of course already well-known \cite{Benzi:2013}; the present study additionally shows, using the generalised q-adjacency kernels, that there is a smooth transition in the q-divergence between the regions corresponding to Estrada's communicability and the Katz centrality.

\clearpage
%


\end{document}